\documentclass[iop]{emulateapj}

\slugcomment{The Astrophysical Journal - Supplement Series 09/2013}

\shorttitle{Kinetic theory for off-equatorial plasma tori}
\shortauthors{Cremaschini et al.}

\begin{document}


\title{Kinetic theory of equilibrium axisymmetric collisionless plasmas\\
in off-equatorial tori around compact objects}


\author{Claudio Cremaschini, Ji\v{r}\'{\i} Kov\'{a}\v{r}, Petr Slan\'{y} and Zden\v{e}k Stuchl\'{\i}k}
\affil{Institute of Physics, Faculty of Philosophy and Science, Silesian University
in Opava, Bezru\v{c}ovo n\'{a}m.13, CZ-74601 Opava, Czech Republic}

\author{Vladim\'{\i}r Karas}
\affil{Astronomical Institute, Academy of Sciences, Bo\v{c}n\'{\i} II, CZ-14131
Prague, Czech Republic}






\begin{abstract}
The possible occurrence of equilibrium off-equatorial tori in the
gravitational and electromagnetic fields of astrophysical compact objects
has been recently proved based on non-ideal MHD theory. These stationary
structures can represent plausible candidates for the modelling of coronal
plasmas expected to arise in association with accretion discs. However,
accretion disc coronae are formed by a highly diluted environment, and so
the fluid description may be inappropriate. The question is posed of whether
similar off-equatorial solutions can be determined also in the case of
collisionless plasmas for which treatment based on kinetic theory, rather
than fluid one, is demanded. In this paper the issue is addressed in the framework of the Vlasov-Maxwell
description for non-relativistic multi-species axisymmetric plasmas subject
to an external dominant spherical gravitational and dipolar magnetic field.
Equilibrium configurations are investigated and explicit solutions for the
species kinetic distribution function are constructed, which are expressed
in terms of generalized Maxwellian functions characterized by isotropic
temperature and non-uniform fluid fields. The conditions for the existence of off-equatorial tori are investigated. It is
proved that these levitating systems are admitted under general conditions
when both gravitational and magnetic fields contribute to shaping the
spatial profiles of equilibrium plasma fluid fields. Then, specifically-kinetic effects carried by the equilibrium solution are
explicitly provided and identified here with diamagnetic, energy-correction
and electrostatic contributions. It is shown that these kinetic\ terms characterize the plasma equation of
state by introducing non-vanishing deviations from the assumption of thermal
pressure.
\end{abstract}


\keywords{accretion discs, off-equatorial tori, collisionless plasmas, kinetic theory}



\section{Introduction}

This paper concerns the Vlasov-Maxwell description of collisionless
magnetized plasmas \citep{book1,book2,book3,book6} related to axisymmetric
discs arising in the combined gravitational and electromagnetic (EM) fields
of astrophysical compact objects. In particular, in this work an application
of the kinetic theory developed by %
\citet{Cr2010,Cr2011a,Cr2011,Cr2012,Cr2013,Cr2013b} is
considered, to the treatment of equilibrium and possibly non-neutral
equatorial as well as off-equatorial plasma tori. Remarkably, the existence of equilibrium levitating (i.e., off-equatorial) structures has been recently pointed out by %
\citet{JK1}, based on a fluid non-ideal magnetohydrodynamics (MHD)
description. Indication of existence of structures of this type occurring in collisionless plasmas is also supported by other recent works based on test-particle approach \citep{ora1,ora2,stu}. In addition, in the context of charge-separated pulsar magnetospheres, \citet{neuk} found an indication of the possible development of stable off-equatorial tori that can be revealed in his early numerical simulations of magnetized collisionless plasmas. Indeed, from the physical point of view, off-equatorial plasma configurations are intrinsically different from the case of equatorial disc
systems. The astrophysical relevance of these structures lies in the
possibility of modelling coronal plasmas characterized by low-density and
high-temperature conditions \citep{corona1,corona2}. In this regard, two different issues arise,
which deserve a detailed investigation. First, under these conditions fluid
descriptions may become inappropriate, requiring in principle the adoption
of a kinetic treatment. However, it still remains to be ascertained whether
the levitating structures can be recovered as equilibrium solutions in the
framework of a kinetic description. Second, \textquotedblleft a
priori\textquotedblright\ it is not obvious whether the kinetic theory
developed previously for equatorial plasmas can be extended to the
description of off-equatorial tori, how this can be achieved in practice and
what are the physical implications as far as their occurrence in real
systems is concerned.

The statistical description of plasma dynamics can be carried out in terms
of either fluid or kinetic treatments, with the choice of the appropriate
framework depending on the plasma phenomenology to be addressed and the
relevant features of the phenomena to be studied \citep{book4,book6}. The
majority of fluid approaches are based on hydrodynamic or MHD treatments %
\citep{book5}. In the case of collisionless plasmas, when these are formulated independently of an underlying
kinetic theory, some limitations can arise. First, it is well known that the
set of fluid equations may not be closed, requiring in principle the
prescription of arbitrary higher-order fluid fields and closure conditions,
including in particular the equation of state (EoS) or the pressure tensor %
\citep{book6}. Second, in these approaches typically no account is given of
microscopic phase-space particle dynamics together with phase-space plasma
collective phenomena. On the other hand, only in the context of kinetic
theory these difficulties can be consistently overcome, as this treatment
permits to obtain well-defined constitutive equations for the relevant fluid
fields describing the plasma state and to solve at the same time the closure
problem \citep{Cr2011}. These issues become relevant in the case of
collisionless or weakly-collisional multi-species plasmas subject to EM and
gravitational fields where phase-space particle dynamics is expected to play
a dominant role. In particular, kinetic theory is essential for studying
both stationary configurations and dynamical evolution of plasmas when
kinetic effects are relevant, such as ones associated with conservation of
particle adiabatic invariants \citep{Cr2013}, temperature and pressure
anisotropies, diamagnetic and finite Larmor-radius (FLR) effects as well as
energy-correction contributions \citep{Cr2010,Cr2011a,Cr2011,Cr2013}.

The problem of formulating a kinetic theory appropriate for the description
of collisionless plasmas in quasi-stationary (i.e., equilibrium)\
configurations of astrophysical accreting systems and laboratory scenarios
has been presented in series of works, based on the non-relativistic
Vlasov-Maxwell description. In our context, several issues have been
treated, ranging from laboratory plasmas occurring in Tokamak devices %
\citep{Cr2011a}, axisymmetric accretion disc plasmas\ characterized by
locally-nested magnetic surfaces \citep{Cr2010,Cr2011,Cr2012}
and current-carrying magnetic loops \citep{Cr2013b} around compact objects
as well as spatially non-symmetric systems in astrophysical and laboratory
contexts \citep{Cr2013}. It was shown that consistent solutions of the
Vlasov equation can be determined for the species kinetic distribution
function (KDF) describing collisionless plasmas, based on the identification
of the relevant single-particle invariants. The equilibrium KDFs were
expressed in terms of generalized bi-Maxwellian distributions, characterized
by temperature anisotropy, non-uniform fluid fields and local plasma flows.
Chapman-Enskog representations of these equilibria were obtained by
developing a suitable perturbative kinetic theory, which in turn made
possible the analytical calculation of the fluid fields and the
identification of the relevant kinetic effects included in the corresponding
MHD description. As a basic consequence, it was shown that these solutions
can exhibit non-vanishing current densities which can also support a kinetic
dynamo mechanism for the self-generation of EM fields in which the plasma is
immersed \citep{Cr2010,Cr2011}. Finally, more recently the
kinetic theory has been extended to describe axisymmetric plasmas
characterized by strong shear-flow and/or supersonic velocities \citep{SS},
while \citet{PRL} reports a kinetic analysis of the stability properties of
particular equilibrium solutions with respect to axisymmetric EM
perturbations.

A notable feature of the kinetic equilibria mentioned here is the unique
prescription of the functional dependences of an appropriate set of fluid
fields, carried by the species KDFs, which are related to physical
observables of the system. These fields are referred to as \textit{structure
functions }and are denoted as $\left\{ \Lambda _{\mathrm{s}}\right\} $ (see
in particular \cite{Cr2010,Cr2011a,Cr2011,Cr2013} and the definition below),
with the subscript \textquotedblleft s\textquotedblright\ being the species
index. Depending on the kinetic regime being considered, according to the
classification scheme presented by \citet{Cr2012}, these dependences are
expressed in terms of the poloidal flux function $\psi $ of the magnetic
field and/or the effective potential $\Phi _{\mathrm{s}}^{\mathrm{eff}}$
defined as%
\begin{equation}
\Phi _{\mathrm{s}}^{\mathrm{eff}}=\Phi +\frac{M_{\mathrm{s}}}{Z_{\mathrm{s}}e%
}\Phi _{\mathrm{G}},  \label{effective}
\end{equation}%
where $\Phi _{\mathrm{G}}$ and $\Phi $ are the gravitational and
electrostatic (ES) potentials respectively, with $M_{\mathrm{s}}$ and $Z_{%
\mathrm{s}}e$ denoting the species particle mass and charge. Thus, in the
general case kinetic theory requires that $\Lambda _{\mathrm{s}}=\Lambda _{%
\mathrm{s}}\left( \psi ,\Phi _{\mathrm{s}}^{\mathrm{eff}}\right) $. It must
be stressed that, behind the apparent simplicity of the result, for
practical applications one ultimately needs obtaining a representation of $%
\Lambda _{\mathrm{s}}$ in terms of spatial coordinates, e.g. cylindrical
ones $\left( R,\varphi ,z\right) $. Assuming an axisymmetric configuration
where $\psi $ and $\Phi _{\mathrm{s}}^{\mathrm{eff}}$ are generic functions
of both $\left( R,z\right) $, the representation $\Lambda _{\mathrm{s}%
}=\Lambda _{\mathrm{s}}\left( \psi ,\Phi _{\mathrm{s}}^{\mathrm{eff}}\right)
=\overline{\Lambda }_{\mathrm{s}}\left( R,z\right) $ applies. Hence, the
complete solution of the problem actually requires determining the explicit
representation of the potentials $\left( \psi ,\Phi _{\mathrm{G}},\Phi
\right) $ in terms of the spatial cylindrical coordinates. This can be a
very difficult task, since in the general case the plasma itself contributes
to the generation of the gravitational filed, through its non-vanishing
mass-density, and, more important, of the EM fields by means of non-vanishing charge and current densities. In the present discussion we ignore,
however, the self-generation of the gravitational field, focusing only on the
generation of EM fields. It follows that, in order to obtain explicitly the
relationship between the EM potentials and the coordinate system, one has
necessarily to solve\ (numerically) the coupled set of Vlasov-Maxwell
equations to determine $\psi =\psi \left( R,z\right) $ and $\Phi =\Phi
\left( R,z\right) $, where the source terms of the fields are prescribed
functions of the potentials. On the other hand, such a solution is also
demanded for the following additional reasons: a) in order to calculate
explicitly the characteristic kinetic effects which enter the equilibrium
KDF and are associated with diamagnetic-FLR and energy-correction effects
(see \cite{Cr2010,Cr2011a,Cr2011,Cr2012,Cr2013} and the discussion in
Section 7); b) in order to establish the diffeomorphism $\mathcal{J}:\left(
R,z\right) \leftrightarrow \left( \psi ,\vartheta \right) $ which relates
cylindrical (and similarly, spherical coordinates) to local magnetic
coordinates, where $\vartheta $ is an angle-like coordinate defined on
equipotential magnetic surfaces $\psi =const.$ The complexity of the theory,
as far as this issue is concerned, might represent a possible limit for the
practical realization of kinetic equilibria of this type for configurations
of astrophysical interest. This may be relevant especially when demanding a
comparison between kinetic and fluid treatments based on analytical
solutions. Therefore, the question arises of whether such a difficulty can
be actually encompassed in some scenarios, possibly by invoking suitable
asymptotic kinetic orderings to be imposed on the system. In particular,
this concerns the existence of configurations which allow for the
construction of the diffeomorphism $\mathcal{J}$ based on analytical
solutions of the potentials $\left( \psi ,\Phi _{\mathrm{G}},\Phi \right) $
in such a way to afford an explicit treatment of the spatial dependences of
the equilibrium fluid fields carried by the KDF according to the constraints
posed by the Vlasov equation.

A second point of crucial importance concerns the determination of the EoS
that characterizes disc plasmas in the collisionless state. As mentioned
above, this is usually realized by prescribing the form of the scalar
pressure, or more generally the components of the pressure tensor, which
represent the closure condition for the Euler momentum equation in MHD
treatments. Since collisionless plasmas are intrinsically characterized by
the occurrence of phase-space anisotropies which cause the equilibrium KDF
to generally deviate from a simple isotropic Maxwellian distribution, the
knowledge of the correct form of the EoS is not a trivial task. In fact, one
can only prescribe the pressure tensor in consistent way on the base of the
kinetic theory (kinetic closure conditions). In this regard, the problem
consists in the identification of the kinetic effects which must be included
in the pressure tensor, the understanding of their physical origin and the
way they influence the macroscopic configuration of the plasma system. Among
the relevant ones, contributions associated with ES corrections can play a
central role, since they arise from microscopic charge interactions and
carry information about local ES fields generated by the validity of
quasi-neutrality condition in rotating plasmas or deviations away from it in
non-neutral systems.

Finally, from the astrophysical point of view a further motivation is
represented by \citet{JK1}, where the discovery of the possible occurrence
of stationary configurations of disc plasmas in off-equatorial tori is
reported. In that work, a Newtonian axisymmetric model of non-conductive,
charged and perfect fluid tori orbiting in the combined gravitational and
dipolar magnetic fields generated by a central compact object is presented.
The result is obtained in the framework of a non-ideal MHD description and
shows that the interplay between gravitational and magnetic fields can effectively
enhance vertically extended structures in a plasma torus, which may
correspond to localized concentrations of matter above and under the
equatorial plane. The importance of this conclusion lies in the possibility
of interpreting these off-equatorial tori as forming the surrounding
material usually invoked to explain the spectral emission/absorption
features in accretion-disc systems. Possible examples of this type are
provided by coronal halos consisting of non-neutral ion-electron plasmas or
by obscuring dusty-plasma tori that are believed to be produced in galactic
nuclei \citep{AGN1,AGN2,AGN3,AGN4,AGN5,AGN6,AGN7,AGN8,AGN9,AGN10}.

The complete understanding of the physical properties of these structures is
far from being satisfactory and deserves further investigations, both
theoretical and observational. In particular, here we consider the
possibility of a low-density and high-temperature coronal plasma for which
the collisionless state applies. Following the arguments discussed above,
under these conditions the proper framework for the description of these
plasmas is represented by the kinetic theory. In particular, the question is
posed of whether kinetic equilibria can be proved to exist for collisionless
magnetized plasmas which exhibit off-equatorial maxima in the matter
distribution and how these solutions can be possibly related to fluid MHD
ones. This amounts at identifying the appropriate kinetic regimes which meet
these conditions and determining the spatial dependences of the
corresponding fluid fields, including the pressure tensor and the
intrinsically-kinetic effects contained in the EoS. To be successful, a
program of this type must make possible an analytical approach in accordance
with the considerations presented above, to extend the kinetic theory
developed in %
\citet{Cr2010,Cr2011a,Cr2011,Cr2012,Cr2013,Cr2013b} and to
permit its practical application to gain insights into an astrophysical
issue of notable importance connected with accretion-disc phenomenology.

To conclude, we should comment on the fact that the reference works cited above on kinetic theory and off-equatorial tori as well as the present investigation are carried out in the framework of a non-relativistic description (both with respect to the treatment of the gravitational field and the plasma velocities). However, the problem of general-relativistic solutions could in principle be posed, since there is a number of situations in which the formation of accretion discs is due to strong gravitational fields, e.g. around black holes, and general-relativistic corrections must be considered. Although a complete theory of this type is still missing, non-relativistic treatments can nevertheless provide the reference framework for the inclusion of some relevant features characteristic of general-relativistic theories. This can be achieved for example by the adoption of pseudo-Newtonian potentials for the description of spherically-symmetric gravitational fields \citep{PW,Z08}. It has been shown, in fact, that the precision of the pseudo-Newtonian description of stationary general-relativistic phenomena can be very high \citep{dt}. An alternative consists in the inclusion of post-Newtonian corrections. In this reference, a kinetic theory of self-gravitating collisionless gases adopting such a technique can be found in \citet{PN1} for spherical solutions, and in \citet{PN2} for the case of axially-symmetric solutions.

\section{Goals and scheme of the paper}

In view of the considerations presented above, the purpose of this paper is
the formulation of a kinetic theory appropriate for the analytical treatment
of collisionless disc plasmas in axisymmetric off-equatorial tori (see Fig. %
\ref{fig-1} for a schematic view of the configuration geometry). The results
of the investigation are the following ones:

\begin{figure}
\epsscale{.90}
\plotone{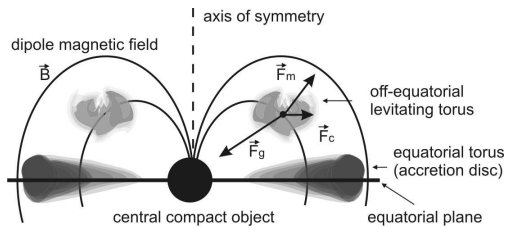}
\caption{Schematic illustration of a poloidal section across the equilibrium axially symmetric distribution of electrically charged matter subject to a dipolar magnetic field and surrounding a gravitating body in the center. Directions of gravitational ($\vec{F}_{\rm g}$), magnetic ($\vec{F}_{\rm m}$) and centrifugal ($\vec{F}_{\rm c}$) forces acting on a generic fluid element belonging to the levitating torus and moving along the circular orbit are displayed.}.\label{fig-1}
\end{figure}


1) The identification of a physically-realizable astrophysical configuration
in which the spatial profiles of the potentials $\left( \psi ,\Phi _{\mathrm{%
s}}^{\mathrm{eff}}\right) $ can be analytically prescribed, when suitable plasma
orderings apply. In the framework of an
asymptotic theory, this permits to decouple the Vlasov equation from the
Maxwell equations, at least to leading-order, with the possibility of an
explicit treatment of the spatial dependences contained in the equilibrium
plasma fluid fields.

2)\ The construction of species equilibrium KDFs which are consistent with
the kinetic constraints imposed by microscopic phase-space conservation laws
for the single particle dynamics. These are proved to be expressed in terms
of generalized Maxwellian KDFs characterized by non-uniform fluid fields and
isotropic temperature, both to be associated with a finite set of structure
functions.

3) The identification of the relevant kinetic regimes which can in principle
arise in the configuration determined at point 1) above and the prescription
of the corresponding functional dependences on the structure functions.

4)\ The development of a perturbative theory which makes possible a
representation of the equilibrium KDFs in terms of a Chapman-Enskog series.
As a result, this permits first the analytical evaluation of the plasma fluid fields, and second to distinguish the characteristic kinetic effects which enter the solution and their main physical properties. 

5) The proof that the gravitationally-bound and magnetized
plasma regime provides the most general functional dependences in the
structure functions carried by the equilibrium KDF, which are consistent
both with the analytical treatment of the fluid fields and the occurrence of
off-equatorial tori. As an illustration of the technique, explicit
calculation of the leading-order species number density and velocity
profiles is provided for two configurations of physical interest.

6) The calculation of the EoS for the equilibrium plasma, to be expressed in terms of the species pressure tensor associated with the KDF. By making use of the perturbative treatment, this includes also the identification of both the kinetic effects and the ES
corrections that can effectively contribute to the EoS. In particular, the latter are shown to determine non-trivial deviations from the assumption of having thermal pressure for the collisionless plasma.

7) To determine the constraints to be imposed on the kinetic solution for
the validity of the theory and the physical configuration realized, which
arise from the Maxwell equations for the self-generated equilibrium EM
fields. These represent necessary conditions which must be verified
\textquotedblleft a posteriori\textquotedblright\ for the complete solution
of the Vlasov-Maxwell problem and the explicit treatment of ES corrections
in comparison with fluid-based approaches.

The scheme of the paper is as follows. In Section 3 the model assumptions
and the fundamental EM orderings are presented. Section 4 deals with the
definition of plasma orderings and the introduction of corresponding kinetic
regimes which characterize collisionless plasmas treated here. In Section 5
the equilibrium species KDFs are explicitly determined and their
realizations are given for three different plasma regimes. In Section 6 a
suitable perturbative theory is developed, which allows for the analytical
treatment of the equilibrium KDFs and the corresponding fluid fields. The
conditions for the occurrence of off-equatorial tori are then investigated
in the case of collisionless plasmas belonging to the gravitationally-bound
and magnetized plasma regime. In Section 7 the expression of the kinetic
corrections which characterize the kinetic equilibria is provided, while the
corresponding contributions in the EoS are computed in
Section 8. Section 9 contains analysis of the Poisson and the Ampere
equations and the constraints which they pose on the kinetic solution.
Finally, concluding remarks are summarized in Section 10.

\section{Model assumptions}
For the construction of kinetic equilibria, we ignore the possible existence
of weakly-dissipative effects (Coulomb collisions and turbulence) and EM
radiation effects \citep{EPJ5,EPJ6}. It is assumed that the
KDF and the EM fields associated with the plasma obey the system of
Vlasov-Maxwell equations, with Maxwell's equations being considered in the
quasi-static approximation. For definiteness, we shall consider here a
plasma consisting of $s$-species of charged particles which are
characterized by proper mass $M_{\mathrm{s}}$ and total charge $Z_{\mathrm{s}%
}e$. In particular, given a generic species KDF $f_{\mathrm{s}}=f_{\mathrm{s}%
}\left( \mathbf{r},\mathbf{v},t\right) $ defined in the phase-space $\Gamma
=\Gamma _{\mathbf{r}}\times \Gamma _{\mathbf{v}}$, with $\Gamma _{\mathbf{r}%
} $ and $\Gamma _{\mathbf{v}}$ being the configuration and velocity space
respectively, the Vlasov equation determines the dynamical evolution of $f_{%
\mathrm{s}}$ and is given by%
\begin{equation}
\frac{\mathrm{d}}{\mathrm{dt}}f_{\mathrm{s}}\left( \mathbf{r},\mathbf{v}%
,t\right) =0.  \label{Vlasov}
\end{equation}

The plasma is taken to be: a) non-relativistic, in the sense that the flow
velocities of all species are small compared to the speed of light $c$, that
the gravitational field can be treated within the classical Newtonian
theory, and that the non-relativistic Vlasov kinetic equation is used as the
dynamical equation for the KDF;\ b) collisionless, so that the mean free
path of the plasma particles is much longer than the largest characteristic
scale length of the plasma; c)\ axisymmetric, so that the relevant dynamical
variables characterizing the plasma (e.g., the fluid fields) are independent
of the azimuthal angle $\varphi ,$ when referred to a set of either
cylindrical coordinates $(R,\varphi ,z)$ or spherical coordinates $%
(r,\varphi ,\theta )$. Thanks to the axisymmetry assumption, as a shortway
in the following we shall denote with $\mathbf{x}$ the configuration state
vector, where $\mathbf{x}$ denotes either $\mathbf{x}=\left( R,z\right) $ or 
$\mathbf{x}=\left( r,\theta \right) $.

We are concerned here with quasi-stationary configurations, namely solutions
which are slowly-varying in time. This condition is also referred to as
equilibrium configuration. For a generic physical quantity $G$ that depends
on spatial coordinates $\mathbf{x}$ and time $t$, the quasi-stationarity is
expressed by letting in the following $G=G\left( \mathbf{x},\lambda
^{k}t\right) $, with $\lambda \ll 1$ being a small dimensionless parameter
to be suitably defined (see below) and $k\geq 1$ an integer. Similar
considerations apply for the equilibrium KDF $f_{\mathrm{s}}$, which is
denoted in the following as $f_{\mathrm{s}}=f_{\mathrm{s}}\left( \mathbf{x},%
\mathbf{v},\lambda ^{k}t\right) $.

From the symmetry properties introduced here one can immediately derive the
fundamental quantities which are conserved for the single-particle dynamics.
In particular, under the assumptions of axisymmetry the canonical momentum
conjugate to the azimuthal angle $\varphi $ is an integral of motion. This
is given by:%
\begin{equation}
P_{\varphi \mathrm{s}}=M_{\mathrm{s}}R\mathbf{v}\cdot \mathbf{e}_{\varphi }+%
\frac{Z_{\mathrm{s}}e}{c}\psi \equiv \frac{Z_{\mathrm{s}}e}{c}\psi _{\ast 
\mathrm{s}}.  \label{p_fi}
\end{equation}%
Furthermore, from the condition of quasi-stationarity, the total particle
energy%
\begin{equation}
E_{\mathrm{s}}=\frac{M_{\mathrm{s}}}{2}v^{2}\mathbf{+}{Z_{\mathrm{s}}e}\Phi
_{\mathrm{s}}^{\mathrm{eff}}(\mathbf{x},\lambda ^{k}t)\equiv Z_{\mathrm{s}%
}e\Phi _{\ast \mathrm{s}},  \label{total_energy}
\end{equation}%
represents an adiabatic invariant of prescribed order, with $\Phi _{\mathrm{s%
}}^{\mathrm{eff}}$ being defined in Eq. (\ref{effective}). Following the
discussion in \citet{Cr2010,Cr2011}, here we recall that a generic quantity $%
P=P\left( \mathbf{x},\mathbf{v},\lambda ^{n}t\right) $ defined in the
phase-space is an adiabatic invariant of order $n$ with respect to $\lambda $
when it satisfies the condition $\frac{1}{\Omega _{\mathrm{cs}}}\frac{%
\mathrm{d}}{\mathrm{dt}}\ln P=0+O(\lambda ^{n+1})$, where $n\geq 0$ is a
suitable integer and $\Omega _{\mathrm{cs}}\equiv \frac{{Z_{\mathrm{s}}eB}}{%
M_{\mathrm{s}}c}$ is the cyclotron frequency. This means that adiabatic
invariants are conserved in asymptotic sense, namely up to a prescribed
order of accuracy determined by the parameters $\lambda $ and $n$.

We consider solutions of the equilibrium magnetic field $\mathbf{B}$ which
admit, at least locally, a family of nested axisymmetric toroidal magnetic
surfaces $\left\{ \psi (\mathbf{x},\lambda ^{k}t)\right\} \equiv \left\{ \psi (\mathbf{x},\lambda ^{k}t%
)=const.\right\} $, where $\psi $ denotes the poloidal magnetic flux of $%
\mathbf{B}$. The magnetic surfaces can be either locally closed %
\citep{Cr2010} or locally open \citep{Cr2011} in the configuration domain
occupied by the plasma. In both cases a set of magnetic coordinates ($\psi
,\varphi ,\vartheta $) can be defined locally, where $\vartheta $ is a
curvilinear angle-like coordinate on the magnetic surfaces $\psi (\mathbf{x}%
,\lambda ^{k}t)=const.$ By construction, magnetic coordinates are related to
cylindrical or spherical coordinates by a diffeomorphism $J$ which must be
consistently determined, as discussed above. Each relevant physical quantity 
$G\left( \mathbf{x},\lambda ^{k}t\right) $ can then be conveniently
expressed either in terms of the set $\mathbf{x}$ or as a function of the
magnetic coordinates, i.e. $G\left( \mathbf{x},\lambda ^{k}t\right) =%
\overline{G}\left( \psi ,\vartheta ,\lambda ^{k}t\right) $.

Consistent with these assumptions, we require the EM field to be
slowly-varying in time, i.e., of the form 
\begin{eqnarray}
\mathbf{E} &=&\mathbf{E}(\mathbf{x},\lambda ^{k}t),  \nonumber \\
\mathbf{B} &=&\mathbf{B}(\mathbf{x},\lambda ^{k}t).  \label{b0}
\end{eqnarray}%
In particular, we assume the magnetic field to be represented as%
\begin{equation}
\mathbf{B}\equiv \nabla \times \mathbf{A}=\mathbf{B}^{\mathrm{self}}(\mathbf{%
x},\lambda ^{k}t)+\mathbf{B}^{\mathrm{ext}}(\mathbf{x},\lambda ^{k}t),
\label{b1}
\end{equation}%
where $\mathbf{B}^{\mathrm{self}}$ and $\mathbf{B}^{\mathrm{ext}}$ denote
the self-generated magnetic field produced by the plasma and a
finite external axisymmetric magnetic field (vacuum field). For
definiteness, in this treatment both contributions are assumed to exhibit
only non-vanishing poloidal components, to be denoted in the following as $\mathbf{B}_{\mathrm{P}}$. Notice that, for what concerns the
self-field, this assumption must be verified \textquotedblleft a
posteriori\textquotedblright\ to be consistent with the constraints placed
on the kinetic solution by the Maxwell equations. Hence, the two fields are
represented as%
\begin{eqnarray}
\mathbf{B}^{\mathrm{ext}} &=&\nabla \psi _{\mathrm{ext}}(\mathbf{x},\lambda
^{k}t)\times \nabla \varphi , \\
\mathbf{B}^{\mathrm{self}} &=&\nabla \psi _{\mathrm{self}}(\mathbf{x}%
,\lambda ^{k}t)\times \nabla \varphi ,  \label{bself}
\end{eqnarray}%
so that the total poloidal magnetic field takes the form%
\begin{equation}
\mathbf{B}\equiv\mathbf{B}_{\mathrm{P}}=\nabla \psi (\mathbf{x},\lambda ^{k}t)\times \nabla \varphi ,
\label{B FIELD}
\end{equation}%
with $%
\psi \equiv \psi _{\mathrm{ext}}+\psi _{\mathrm{self}}$. In particular, for
the purpose of the present work, the external magnetic field is taken to
coincide with a dipolar field. In such a case, the flux function $\psi
_{\mathrm{ext}} $ is written in terms of the spherical coordinates $\left( r,\theta
\right) $ as%
\begin{equation}
\psi _{\mathrm{ext}}=\mathcal{M}_{\mathrm{0}}\frac{\sin ^{2}\theta }{r},
\label{dipolo}
\end{equation}%
with $\mathcal{M}_{\mathrm{0}}$ being the magnitude of the dipole magnetic
moment.

Charged particles are also assumed to be subject to the effective potential $%
\Phi _{\mathrm{s}}^{\mathrm{eff}}\equiv \Phi _{\mathrm{s}}^{\mathrm{eff}}(%
\mathbf{x},\lambda ^{k}t)$ defined by Eq. (\ref{effective}). In principle,
both the ES potential $\Phi (\mathbf{x},\lambda ^{k}t)$ and the
gravitational potential $\Phi _{\mathrm{G}}(\mathbf{x},\lambda ^{k}t)$ can
be produced by the plasma itself and by external sources. However, in the
following it is assumed that $\Phi (\mathbf{x},\lambda ^{k}t)$ is uniquely
generated by the plasma charge density, while we shall neglect the self
contribution of the plasma to $\Phi _{\mathrm{G}}$. Hence, for an
axisymmetric disc, the gravitational potential is taken as being stationary
and to coincide identically with the potential generated by the central
compact object. The latter is expressed here in terms of the
spherically-symmetric Newtonian potential as%
\begin{equation}
\Phi _{\mathrm{G}}(\mathbf{x})=-\frac{G_{\mathrm{N}}M_{\mathrm{+}}}{r},
\label{Newton}
\end{equation}%
where $G_{\mathrm{N}}$ is the Newton gravitational constant and $M_{\mathrm{+%
}}$ is the mass of the compact object.

Given validity of these assumptions, in order to address the first issue
posed in the Section 2 we proceed introducing the following fundamental
orderings for the self EM fields:\newline
1)\ The self component of the equilibrium magnetic field $\mathbf{B}^{%
\mathrm{self}}$ is ordered with respect to the external field $\mathbf{B}^{%
\mathrm{ext}}$ as%
\begin{equation}
\frac{\left\vert \mathbf{B}^{\mathrm{self}}\right\vert }{\left\vert \mathbf{B%
}^{\mathrm{ext}}\right\vert }\sim O\left( \lambda ^{j}\right) ,
\label{basic}
\end{equation}%
with $j\geq 1$, and in general $j\neq k$.\newline
2) The equilibrium ES potential $\Phi $ satisfies the ordering assumption%
\begin{equation}
\left\vert \frac{Z_{\mathrm{s}}e\Phi }{M_{\mathrm{s}}\Phi _{\mathrm{G}}}%
\right\vert \sim O\left( \lambda ^{j}\right) ,  \label{eees}
\end{equation}%
with $j\geq 1$. This means that the ES potential energy is small with
respect to the gravitational potential energy, where $\lambda $ will be
properly defined below. In the following the case $j=1$ in Eqs. (\ref{basic}%
) and (\ref{eees}) will be considered.

It is important to remark that the two conditions (\ref{basic}) and (\ref%
{eees}) pose strong constraints from the physical point of view on the
realizability of the equilibrium configuration for the collisionless disc
plasma. In particular, they must be verified \textquotedblleft a
posteriori\textquotedblright\ in order to warrant the consistency of the
kinetic equilibrium solution for the species KDF with the validity of the
Maxwell equations.

Eqs. (\ref{basic}) and (\ref{eees}) are the fundamental EM orderings which
permit the analytical treatment of the spatial dependences of the
equilibrium fluid fields as prescribed by the kinetic solution to be
determined below. In fact, when these orderings hold, to leading-order the
magnetic flux function $\psi $ and the effective potential $\Phi _{\mathrm{s}%
}^{\mathrm{eff}}$ coincide respectively with the vacuum fields, namely the
external dipolar flux function $\psi _{\mathrm{ext}}$ in Eq. (\ref{dipolo})
and with the gravitational potential $\Phi _{\mathrm{G}}$ given by Eq. (\ref%
{Newton}). Hence, at this order it is possible to construct explicitly the
diffeomorphism which relates the spherical coordinates $\left( r,\theta
\right) $ with the potentials $\left( \psi ,\Phi _{\mathrm{s}}^{\mathrm{eff}%
}\right) \cong \left( \psi _{\mathrm{ext}},\Phi _{\mathrm{G}}\right) $. This
is expressed by letting%
\begin{eqnarray}
r &=&\left\vert \frac{G_{\mathrm{N}}M_{\mathrm{+}}}{\Phi _{\mathrm{G}}}%
\right\vert ,  \label{d1} \\
\sin ^{2}\theta &=&\left\vert \frac{G_{\mathrm{N}}M_{\mathrm{+}}}{%
\mathcal{M}_{\mathrm{0}}}\frac{\psi _{\mathrm{ext}}}{\Phi _{\mathrm{G}}}%
\right\vert .  \label{d2}
\end{eqnarray}%
The corresponding relationship holding for cylindrical coordinates can then
be obtained from the spherical ones, giving in particular%
\begin{equation}
R=r\sin \theta =\left\vert \frac{G_{\mathrm{N}}M_{\mathrm{+}}}{\Phi _{%
\mathrm{G}}}\right\vert \sqrt{\left\vert \frac{G_{\mathrm{N}}M_{\mathrm{+}}}{%
\mathcal{M}_{\mathrm{0}}}\frac{\psi _{\mathrm{ext}}}{\Phi _{\mathrm{G}}}%
\right\vert }.  \label{tre}
\end{equation}

\section{Plasma orderings and kinetic regimes}

In this section we introduce the relevant orderings for collisionless
plasmas, which permit to identify corresponding kinetic regimes
characterized by different equilibrium solutions and to verify their
consistency with the existence of off-equatorial tori. To this aim we
determine a classification scheme based on the method outlined in %
\citet{Cr2012} and suitable for the treatment of the physical setting
indicated in the previous section.

The first step is the definition of the dimensionless species-dependent
parameters $\varepsilon _{\mathrm{M,s}}$, $\varepsilon _{\mathrm{s}}$ and $%
\sigma _{\mathrm{s}}$. These are prescribed in such a way to be all
independent of single-particle velocity and at the same time to be related
to the characteristic species thermal velocities. Both perpendicular and
parallel thermal velocities (defined with respect to the magnetic field
local direction) must be considered. These are defined respectively by $%
v_{\perp \mathrm{ths}}=\left\{ T_{\perp \mathrm{s}}/M_{\mathrm{s}}\right\}
^{1/2}$\ and $v_{\parallel \mathrm{ths}}=\left\{ T_{\parallel \mathrm{s}}/M_{%
\mathrm{s}}\right\} ^{1/2}$, with $T_{\perp \mathrm{s}}$\ and $T_{\parallel 
\mathrm{s}}$\ denoting here the species perpendicular and parallel
temperatures. In detail, the first parameter is defined as $\varepsilon _{%
\mathrm{M,s}}\equiv \frac{r_{\mathrm{Ls}}}{L}$, where $r_{\mathrm{Ls}%
}=v_{\perp \mathrm{ths}}/\Omega _{\mathrm{cs}}$\ is the species Larmor
radius, with $L$\ being the characteristic scale-length of the spatial
variations of all of the fluid fields associated with the KDF and of the EM
fields. The second parameter $\varepsilon _{\mathrm{s}}$\ is related to the
particle canonical momentum $P_{\varphi \mathrm{s}}$. By denoting $v_{%
\mathrm{ths}}\equiv \sup \left\{ v_{\parallel \mathrm{ths}},v_{\perp \mathrm{%
ths}}\right\} $, $\varepsilon _{\mathrm{s}}$\ is identified with $%
\varepsilon _{\mathrm{s}}\equiv \left\vert \frac{M_{\mathrm{s}}Rv_{\mathrm{%
ths}}}{\frac{Z_{\mathrm{s}}e}{c}\psi }\right\vert $. Hence, $\varepsilon _{%
\mathrm{s}}$\ effectively measures the ratio between the toroidal angular
momentum $L_{\varphi \mathrm{s}}\equiv M_{\mathrm{s}}Rv_{\varphi }$\ and the
magnetic contribution to the toroidal canonical momentum, for all particles
in which $v_{\varphi }$\ is of the order $v_{\varphi }\sim v_{\mathrm{ths}}$%
\ while $\psi $\ is assumed as being non-vanishing. In particular, here the
magnetic flux can be estimated as $\psi \sim B_{\mathrm{p}}RL_{\mathrm{1}}$,
with $L_{\mathrm{1}}$\ denoting the characteristic length-scale of flux
variations and $B_{\mathrm{p}}$ the magnitude of the poloidal magnetic
field. Note that, by definition, $L\leq L_{\mathrm{1}}$, but, in principle, we can also have $L \ll L_{\mathrm{1}}$ locally.\ 

We also notice that the present definition of $\varepsilon _{\mathrm{s}}$ is
not the only possibility, as one could also define the parameter $%
\varepsilon _{\mathrm{s}}$ such that it is related to the azimuthal flow
velocity, namely letting $v_{\varphi }\sim V_{\varphi \mathrm{s}}=\Omega _{%
\mathrm{s}}R$, with $\Omega _{\mathrm{s}}$ being the corresponding angular
frequency. Finally, $\sigma _{\mathrm{s}}$\ is related to the particle total
energy $E_{\mathrm{s}}$\ and is prescribed in this work as $\sigma _{\mathrm{%
s}}\equiv \left\vert \frac{\frac{M_{\mathrm{s}}}{2}v_{\mathrm{ths}}^{2}+Z_{%
\mathrm{s}}e\Phi }{M_{\mathrm{s}}\Phi _{\mathrm{G}}}\right\vert $. It
follows that $\sigma _{\mathrm{s}}$\ measures the ratio between particle
kinetic and ES potential energy with respect to the gravitational potential
energy, for all particles having velocity $v$\ of the order $v\sim v_{%
\mathrm{ths}}$, with $\Phi _{\mathrm{G}}$ being assumed as non-vanishing. We
notice that the definition of $\sigma _{\mathrm{s}}$ differs from that used
in previous works \citep{Cr2011a,Cr2011,Cr2012} while it is consistent with
the ordering assumption (\ref{eees}). In the following we shall denote as
thermal subset of velocity space the subset of the Euclidean velocity space
in which the asymptotic conditions $\frac{v}{v_{\mathrm{ths}}}\sim \frac{%
v_{\varphi }}{v_{\mathrm{ths}}}\sim O\left( 1\right) $ holds.

A comment is in order regarding the role of the magnetic field in the two
parameters $\varepsilon _{\mathrm{s}}$ and $\varepsilon _{\mathrm{M,s}}$. In
the first case, the magnetic field is represented by means of the poloidal
flux $\psi $ which contributes to the toroidal canonical momentum $%
P_{\varphi \mathrm{s}}$, while $\varepsilon _{\mathrm{M,s}}$ depends on the
magnitude of the total magnetic field. Invoking the definitions for $\varepsilon _{\mathrm{s}}$ and $\varepsilon _{\mathrm{M,s}}$\ given above, it
follows that $\varepsilon _{\mathrm{s}}\sim \varepsilon _{\mathrm{M,s}}\frac{%
L}{L_{\mathrm{1}}}\frac{B}{B_{\mathrm{p}}}$, where $L$\ and $L_{\mathrm{1}}$%
\ are respectively the characteristic scale-lengths of equilibrium fluid and
EM fields and of the poloidal flux. In general, the two quantities should be
considered as independent, with $L\leq L_{\mathrm{1}}$ and $B_{\mathrm{p}%
}\leq B$ (in the present context one has identically $B_{\mathrm{p}}=B$\
from Eq. (\ref{B FIELD})). Indeed, the parameter $\varepsilon _{\mathrm{s}}$
determines the particle spatial excursions from a magnetic flux surface,
while $\varepsilon _{\mathrm{M,s}}$\ measures the amplitude of the Larmor
radius with respect to the inhomogeneities of the background fluid fields.
These two effects correspond to different physical magnetic-related
processes, due respectively to the Larmor-radius and magnetic-flux surface
confinement mechanisms.

In this work we assume that the ordering condition%
\begin{equation}
\varepsilon _{\mathrm{M,s}}\ll 1  \label{eps-emme}
\end{equation}%
holds for the collisionless plasma considered here. This amounts at
requiring that the Larmor radius remains small with respect to the
scale-length $L$, which, as shown in \citet{Cr2012}, represents a condition
that is expected to be easily verified in accretion-disc systems. Hence,
one can consistently identify the small parameter $\lambda $ introduced
above with $\lambda =\sup \left\{ \varepsilon _{\mathrm{M,s}}\right\} $.

The classification that is introduced in this work is based on the magnitude
of the two parameters $\varepsilon _{\mathrm{s}}$ and $\sigma _{\mathrm{s}}$%
. In detail, plasma species will be distinguished as belonging to the
following regimes:\newline
1) \emph{Gravitationally-bound }if $\sigma _{\mathrm{s}}\ll 1$ and $%
\varepsilon _{\mathrm{s}}\gtrsim 1$.\newline
2)\ \emph{Magnetized} if $\varepsilon _{\mathrm{s}}\ll 1$ and $\sigma _{%
\mathrm{s}}\lesssim 1$.\newline
3) \emph{Gravitationally-bound and magnetized }if both $\sigma _{\mathrm{s}%
}\ll 1$ and $\varepsilon _{\mathrm{s}}\ll 1$.

In the case of regimes 1 and 3 the following asymptotic expansion holds
for the total particle energy $Z_{\mathrm{s}}e\Phi _{\ast \mathrm{s}}$:%
\begin{equation}
\Phi _{\ast \mathrm{s}}\mathbf{=}\frac{M_{\mathrm{s}}}{Z_{\mathrm{s}}e}\Phi
_{\mathrm{G}}\left[ 1+O\left( \sigma _{\mathrm{s}}\right) \right] .
\label{order-sigma}
\end{equation}%
Similarly, for regimes 2 and 3 the particle canonical momentum $\frac{Z_{\mathrm{s}}e}{c}\psi
_{\ast \mathrm{s}}$ admits the expansion%
\begin{equation}
\psi _{\ast \mathrm{s}}=\psi \left[ 1+O\left( \varepsilon _{\mathrm{s}%
}\right) \right] .  \label{ordering-epsilon}
\end{equation}%
It is instructive to analyze the main features of these regimes and the
physical conditions for their occurrence. The action of some energy
non-conserving mechanisms is required for the establishment of the case of
gravitationally-bound plasmas. In particular, plausible physical mechanisms
that can be responsible for the decrease of the single-particle kinetic
energy, in both collisionless and collisional AD plasmas, are EM
interactions (e.g., binary Coulomb collisions among particles and
particle-wave interactions, such as Landau damping) and/or radiation
emission (radiation-reaction). These can in principle be ascribed also to
the occurrence of EM instabilities and EM turbulence. For single particles
these processes can be dissipative. As a consequence, these particles tend
to move towards regions with higher gravitational potential (in absolute
value). After multiple interactions of this type, the process can ultimately
reach an equilibrium state which corresponds to the gravitationally-bound
regimes. As far as the magnetic-field based classification, we notice that
the requirement $\varepsilon _{\mathrm{s}}\ll 1$ (regimes 2 and 3) means
that a particle trajectory remains close to the same magnetic surface $\psi
=const.$, while satisfying the ordering (\ref{eps-emme}).

Finally, for greater generality, in the rest of the treatment we shall
assume that, in the regimes in which $\sigma _{\mathrm{s}}\ll 1$ and/or $%
\varepsilon _{\mathrm{s}}\ll 1$, the orderings $\sigma _{\mathrm{s}}\sim
\varepsilon _{\mathrm{M,s}}$ and $\varepsilon _{\mathrm{s}}\sim \varepsilon
_{\mathrm{M,s}}$ apply.

\section{Equilibrium species KDF}

In this section we proceed with the construction of the species equilibrium
KDF and its characterization to the plasma regimes identified above. We
consider both exact as well as asymptotic representations for the solution,
the latter being expressed in terms of a Chapman-Enskog series. To reach the
goal, here we adopt the solution technique developed in %
\citet{Cr2010,Cr2011a,Cr2011,Cr2013,Cr2013b}, which consists in the
construction of solutions of the Vlasov equation of the form $f_{\mathrm{s}%
}=f_{\ast \mathrm{s}}$, where $f_{\ast \mathrm{s}}$\ is a suitable adiabatic
invariant. This amounts at requiring that $f_{\ast \mathrm{s}}$ is
necessarily a function of particle adiabatic invariants. In view of the
model assumptions introduced above, it follows that the general form of the
equilibrium KDF in the present context is given by%
\begin{equation}
f_{\ast \mathrm{s}}=f_{\ast \mathrm{s}}\left( E_{\mathrm{s}},P_{\varphi 
\mathrm{s}},\Lambda _{\ast \mathrm{s}},\lambda ^{k}t\right) ,  \label{fss}
\end{equation}%
with $k\geq 1$ and where slow-time dependences are assumed to be uniquely associated with the particle energy. Here $\Lambda _{\ast \mathrm{s}}$ denotes the so-called
structure functions, i.e., functions which depend implicitly on the particle
state $\left( \mathbf{x},\mathbf{v}\right) $. In order for $f_{\ast \mathrm{s%
}} $ to be an adiabatic invariant, $\Lambda _{\ast \mathrm{s}}$ must also be
functions of the adiabatic invariants. This restriction is referred to here
as a kinetic constraint. The precise form of the functional dependences of $%
\Lambda _{\ast \mathrm{s}}$ is characteristic of each plasma regime, as
discussed below.

In order to determine an explicit representation of $f_{\ast \mathrm{s}}$
according to Eq. (\ref{fss}), we impose the following requirements:

1) The KDF must be characterized by species-dependent non-uniform fluid
fields, azimuthal flow velocity and isotropic temperature, to be suitably
prescribed in terms of the structure functions.

2) Open, locally nested magnetic flux surfaces: the magnetic field is taken
to allow quasi-stationary solutions with magnetic flux lines belonging to
locally nested and generally open magnetic surfaces.

3) Kinetic constraints: suitable functional dependences are imposed on the
structure functions $\Lambda _{\ast \mathrm{s}}$ which depend on the regime
being considered and such to warrant $f_{\ast \mathrm{s}}$ to be an
adiabatic invariant.

4) In all regimes, $f_{\ast \mathrm{s}}$\ is required to be asymptotically
\textquotedblleft close\textquotedblright\ (in a suitable sense to be
defined below)\ to a local Maxwellian KDF. This requires the possibility of
determining \textquotedblleft a posteriori\textquotedblright\ a perturbative
representation of the KDF equivalent to the Chapman-Enskog expansion for the
analytical treatment of implicit phase-space dependences contained in the
structure functions, with the consistent inclusion of ES corrections,
FLR-diamagnetic and/or energy-corrections contributions.

5)\ The KDF $f_{\ast \mathrm{s}}$ must be a strictly-positive real function
and it must be summable, in the sense that the velocity moments of the form%
\begin{equation}
\Xi _{\mathrm{s}}(\mathbf{x},\lambda ^{k}t)=\int_{\Gamma _{\mathbf{v}}}%
\mathrm{{d}\mathbf{v}}K_{\mathrm{s}}(\mathbf{x},\mathbf{v},\lambda ^{k}t)f_{\ast \mathrm{s}}
\label{velomo}
\end{equation}%
must exist for a suitable ensemble of weight functions $\left\{ K_{\mathrm{s}%
}(\mathbf{x},\mathbf{v},\lambda ^{k}t)\right\} $, to be prescribed in terms of
polynomials of arbitrary degree defined with respect to components of the
velocity vector field $\mathbf{v}$.

Then, following \citet{Cr2010,Cr2011a,Cr2011,Cr2013,Cr2013b}, we express the
equilibrium KDF $f_{\ast \mathrm{s}}$ as%
\begin{equation}
f_{\ast \mathrm{s}}=\frac{\eta _{\ast \mathrm{s}}}{\left( 2\pi /M_{\mathrm{s}%
}\right) ^{3/2}T_{\ast \mathrm{s}}^{3/2}}\exp \left\{ -\frac{E_{\mathrm{s}%
}-\Omega _{\ast \mathrm{s}}P_{\varphi \mathrm{s}}}{T_{\ast \mathrm{s}}}%
\right\} ,  \label{gM}
\end{equation}%
which is referred to as the \textit{Generalized Maxwellian KDF.} Here the
structure functions are identified with the set $\Lambda _{\ast \mathrm{s}%
}\equiv \left( \eta _{\ast \mathrm{s}},T_{\ast \mathrm{s}},\Omega _{\ast 
\mathrm{s}}\right) $, where $\eta _{\ast \mathrm{s}}$, $T_{\ast \mathrm{s}}$
and $\Omega _{\ast \mathrm{s}}$ are related to the species number density,
isotropic temperature and azimuthal angular velocity respectively. Invoking
the definitions (\ref{p_fi}) and (\ref{total_energy}), Eq. (\ref{gM}) can
also be written as%
\begin{equation}
f_{\ast \mathrm{s}}=\frac{\eta _{\ast \mathrm{s}}\exp \left[ \frac{X_{\ast 
\mathrm{s}}}{T_{\ast \mathrm{s}}}\right] }{\left( 2\pi /M_{\mathrm{s}%
}\right) ^{3/2}T_{\ast \mathrm{s}}^{3/2}}\exp \left\{ -\frac{M_{\mathrm{s}%
}\left( \mathbf{v}-\mathbf{V}_{\ast \mathrm{s}}\right) ^{2}}{2T_{\ast 
\mathrm{s}}}\right\} ,
\end{equation}%
where $\mathbf{V}_{\ast \mathrm{s}}=R\Omega _{\ast \mathrm{s}}\mathbf{e}%
_{\varphi }$ and%
\begin{equation}
X_{\ast \mathrm{s}}\equiv M_{\mathrm{s}}\frac{\left\vert \mathbf{V}_{\ast 
\mathrm{s}}\right\vert ^{2}}{2}+\frac{Z_{\mathrm{s}}e}{c}\psi \Omega _{\ast 
\mathrm{s}}-Z_{\mathrm{s}}e\Phi _{s}^{\mathrm{eff}}.
\end{equation}

It is worth pointing out that the form of the solution (\ref{gM}) holds for
all the plasma kinetic regimes identified in the previous section. The
difference in the three cases concerns the prescription of the kinetic
constraints to be imposed on $\Lambda _{\ast \mathrm{s}}$. In particular,
consistent with the requirements listed above, these are assigned as follows:

1)\ For gravitationally-bound plasmas it is required that the functional
dependence of $\Lambda _{\ast \mathrm{s}}$ is of the type%
\begin{equation}
\Lambda _{\ast \mathrm{s}}\equiv \Lambda _{\ast \mathrm{s}}\left( \Phi
_{\ast \mathrm{s}}\right) ,  \label{GB-K}
\end{equation}%
for which Eq. (\ref{order-sigma}) applies, while implicit dependences with
respect to $\psi _{\ast \mathrm{s}}$ remain excluded in such a case.

2) For magnetized plasmas, the kinetic constraint is realized by imposing%
\begin{equation}
\Lambda _{\ast \mathrm{s}}\equiv \Lambda _{\ast \mathrm{s}}\left( \psi
_{\ast \mathrm{s}}\right) ,  \label{M-K}
\end{equation}%
for which Eq. (\ref{ordering-epsilon}) applies, while implicit dependences
with respect to $\Phi _{\ast \mathrm{s}}$ are excluded.

3) For gravitationally-bound and magnetized plasmas both Eqs. (\ref%
{order-sigma}) and (\ref{ordering-epsilon}) hold, so that the general form
of the kinetic constraint is given by%
\begin{equation}
\Lambda _{\ast \mathrm{s}}\equiv \Lambda _{\ast \mathrm{s}}\left( \psi
_{\ast \mathrm{s}},\Phi _{\ast \mathrm{s}}\right) .  \label{GB-M-K}
\end{equation}
The connection between the realization of these regimes and the occurrence
of off-equatorial tori will be investigated in the next section. Here it
must be noticed that, because of the constraints (\ref{GB-K})$-$(\ref{GB-M-K}%
), at this stage the structure functions cannot be regarded as fluid fields,
since they are defined in the phase-space, namely they depend on the single
particle velocity via the particle energy $E_{\mathrm{s}}$ and the canonical
momentum $P_{\varphi \mathrm{s}}$. Instead, the fluid fields associated with 
$f_{\ast \mathrm{s}}$ must be properly computed as velocity moments
according to Eq. (\ref{velomo}) and they are unique once the precise form of
the structure functions is explicitly prescribed in $f_{\ast \mathrm{s}}$.

\section{Off-equatorial tori:\ density and velocity profiles}

In this section we first proceed determining a Chapman-Enskog representation
for $f_{\ast \mathrm{s}}$ that makes possible the treatment of the implicit
phase-space functional dependences carried by the structure functions as
well as the analytical evaluation of the equilibrium fluid fields and the
associated kinetic contributions. We then apply the result to prove the
validity of the kinetic theory developed here as far as the description of
off-equatorial toroidal structures is concerned. This task can be achieved
by implementing an appropriate perturbative theory for $f_{\ast \mathrm{s}}$
which was first developed in \citet{Cr2010,Cr2011} and which is based on a
Taylor expansion of $\Lambda _{\ast \mathrm{s}}$ with respect to the
dimensionless parameters $\sigma _{\mathrm{s}}$ and $\varepsilon _{\mathrm{s}%
}$.

It is understood that the basic feature of such a kinetic perturbative
technique is that it is strictly applicable only in localized subsets of
velocity space (thermal subsets), namely to particles whose velocity
satisfies the asymptotic ordering (\ref{order-sigma}) and/or (\ref%
{ordering-epsilon}). A notable consequence of such an approach is that, for
each kinetic regime, quasi-stationary, self-consistent, asymptotic solutions
of the Vlasov-Maxwell equations (kinetic equilibria) can be explicitly
determined by means of suitable Taylor expansions of $f_{\ast \mathrm{s}}$.
In particular, it is found that Maxwellian-like KDFs can be obtained locally
in phase-space, where the appropriate convergence conditions hold. This
procedure provides also the correct constitutive equations of the
leading-order fluid fields as well as the precise form of the ES,
FLR-diamagnetic and/or energy-correction contributions to the KDF.

In detail, invoking Eqs. (\ref{order-sigma}) and (\ref{ordering-epsilon}), a
linear asymptotic expansion for the structure functions can be obtained. In
the general case, neglecting corrections of $O\left( \varepsilon _{\mathrm{s}%
}\sigma _{\mathrm{s}}\right) ,$ as well as of $O\left( \varepsilon _{\mathrm{%
s}}^{k}\right) $ and $O\left( \sigma _{\mathrm{s}}^{k}\right) $, with $k\geq
2$, this is given by%
\begin{eqnarray}
\Lambda _{\ast \mathrm{s}} &\cong &\Lambda _{\mathrm{s}}+\left( \psi _{\ast 
\mathrm{s}}-\psi \right) \left[ \frac{\partial \Lambda _{\ast \mathrm{s}}}{%
\partial \psi _{\ast \mathrm{s}}}\right] _{\psi _{\ast \mathrm{s}}=\psi
,\Phi _{\ast \mathrm{s}}=\frac{M_{\mathrm{s}}}{Z_{\mathrm{s}}e}\Phi _{%
\mathrm{G}}}  \nonumber \\
&&+\left( \Phi _{\ast \mathrm{s}}-\frac{M_{\mathrm{s}}}{Z_{\mathrm{s}}e}\Phi
_{\mathrm{G}}\right) \left[ \frac{\partial \Lambda _{\ast \mathrm{s}}}{%
\partial \Phi _{\ast \mathrm{s}}}\right] _{\psi _{\ast \mathrm{s}}=\psi
,\Phi _{\ast \mathrm{s}}=\frac{M_{\mathrm{s}}}{Z_{\mathrm{s}}e}\Phi _{%
\mathrm{G}}},  \label{qq}
\end{eqnarray}%
where $\Lambda _{\mathrm{s}}$ is the leading-order term which uniquely
follows for each regime (see below). When Eq. (\ref{qq}) is applied to the
equilibrium KDF $f_{\ast \mathrm{s}}$ and the ordering (\ref{eees}) is also
invoked, the following Chapman-Enskog representation is found:%
\begin{equation}
f_{\ast \mathrm{s}}=f_{\mathrm{M,s}}\left[ 1+\varepsilon _{\mathrm{s}}h_{%
\mathrm{s}}^{\left( 1\right) }+\sigma _{\mathrm{s}}h_{\mathrm{s}}^{\left(
2\right) }+\lambda h_{\mathrm{s}}^{\left( 3\right) }\right] ,  \label{CE}
\end{equation}%
where the leading-order contribution $f_{\mathrm{M,s}}$ coincides with a
drifted isotropic Maxwellian KDF carrying non-uniform number density,
azimuthal differential flow velocity and isotropic temperature. In detail:%
\begin{equation}
f_{\mathrm{M,s}}=\frac{n_{\mathrm{s}}}{\left( 2\pi /M_{\mathrm{s}}\right)
^{3/2}T_{\mathrm{s}}^{3/2}}\exp \left\{ -\frac{M_{\mathrm{s}}\left( \mathbf{v%
}-\mathbf{V}_{\mathrm{s}}\right) ^{2}}{2T_{\mathrm{s}}}\right\} ,  \label{fm}
\end{equation}%
where $\mathbf{V}_{\mathrm{s}}=R\Omega _{\mathrm{s}}\mathbf{e}_{\varphi }$
is the leading-order drift velocity carried by $f_{\mathrm{M,s}}$. Here $n_{%
\mathrm{s}}$ represents the leading-order species number density and is
given by%
\begin{equation}
n_{\mathrm{s}}\equiv \eta _{\mathrm{s}}\exp \left[ \frac{\frac{M_{\mathrm{s}}%
}{2}R^{2}\Omega _{\mathrm{s}}^{2}+\frac{Z_{\mathrm{s}}e}{c}\psi \Omega _{%
\mathrm{s}}-M_{\mathrm{s}}\Phi _{\mathrm{G}}}{T_{\mathrm{s}}}\right] ,
\label{nde}
\end{equation}%
with $\eta _{\mathrm{s}}$ being referred to as the pseudo-density. The
leading-order structure functions $\Lambda _{\mathrm{s}}$ coincide now with
the set of functions $\Lambda _{\mathrm{s}}\equiv \left( \eta _{\mathrm{s}%
},T_{\mathrm{s}},\Omega _{\mathrm{s}}\right) $ which are defined in the
configuration space, with $T_{\mathrm{s}}$ and $\Omega _{\mathrm{s}}$ being
respectively the leading-order species temperature and azimuthal rotation
angular frequency. In addition, the quantities $h_{\mathrm{s}}^{\left(
1\right) }$, $h_{\mathrm{s}}^{\left( 2\right) }$ and $h_{\mathrm{s}}^{\left(
3\right) }$ represent first-order kinetic corrections. In particular, $h_{%
\mathrm{s}}^{\left( 1\right) }$ is referred to as FLR-diamagnetic
contribution, $h_{\mathrm{s}}^{\left( 2\right) }$ carries energy-correction
contributions (with respect to both kinetic and ES potential energies),
while $h_{\mathrm{s}}^{\left( 3\right) }$ represents a purely ES term.

The precise expression of these functions will be given below in a separate
section, where we discuss the relevance of these kinetic effects and their
physical meaning in the framework of the present perturbative theory. For
the moment, it is sufficient to say that all the first-order corrections are
part of the kinetic equilibrium, and cannot be neglected for the consistent
formulation of the solution. It must be also stressed here that Eq. (\ref{CE}%
) is very general: while $h_{\mathrm{s}}^{\left( 3\right) }$ is
non-vanishing for all the kinetic regimes considered above, the existence of 
$h_{\mathrm{s}}^{\left( 1\right) }$ and $h_{\mathrm{s}}^{\left( 2\right) }$
depends instead on the type of kinetic constraints. In particular, we
distinguish the following features:

1)\ For gravitationally-bound plasmas (regime 1) $h_{\mathrm{s}}^{\left(
1\right) }=0$ and $\Lambda _{\mathrm{s}}$ is subject to the constraint%
\begin{equation}
\Lambda _{\mathrm{s}}=\Lambda _{\mathrm{s}}\left( \Phi _{\mathrm{G}}\right) .
\label{R1}
\end{equation}

2) For magnetized plasmas (regime 2) $h_{\mathrm{s}}^{\left( 2\right) }=0$
and the functional dependence of $\Lambda _{\mathrm{s}}$ becomes%
\begin{equation}
\Lambda _{\mathrm{s}}=\Lambda _{\mathrm{s}}\left( \psi \right) .  \label{R2}
\end{equation}

3) For gravitationally-bound and magnetized plasmas (regime 3) in general
both $h_{\mathrm{s}}^{\left( 1\right) }\neq 0$ and $h_{\mathrm{s}}^{\left(
2\right) }\neq 0$, while for $\Lambda _{\mathrm{s}}$ one has in this case%
\begin{equation}
\Lambda _{\mathrm{s}}=\Lambda _{\mathrm{s}}\left( \psi ,\Phi _{\mathrm{G}%
}\right) .  \label{R3}
\end{equation}

We notice that, to the leading-order, the equilibrium solution determined
here does not depend on the ES potential, but only on the gravitational
potential $\Phi _{\mathrm{G}}$ and the magnetic flux $\psi \cong \psi _{%
\mathrm{ext}}$, which are assigned and known functions of the spatial
coordinates.

The perturbative theory developed here represents the starting point for the
application of the kinetic theory to the modelling of off-equatorial plasma
tori in axisymmetric disc systems. This is based on the analysis of the
spatial dependences which characterize the leading-order kinetic solution
and that can be dealt with analytically thanks to the fundamental EM
ordering assumptions introduced in Section 3. In particular, for each of the
three regimes considered here, the proof that the kinetic equilibria admit
off-equatorial solutions follows by analyzing the spatial profile of the
leading-order number density defined by Eq. (\ref{nde}) under the
requirement of having maxima out of the equatorial plane, namely for $\theta
\neq \frac{\pi }{2}$. This requires preliminary to assign the functional
form of the structure functions $\Lambda _{\mathrm{s}}$ characterizing Eq. (%
\ref{nde}) in terms of the potentials $\psi $ and/or $\Phi _{\mathrm{G}}$. A
detailed discussion of this type for all of the three plasma regimes is
beyond the scope of this work and will be addressed separately in future
studies. For the purpose of the present investigation, it is sufficient to
consider here the case of regime 3. In fact, this provides the most general
conditions for the occurrence of levitating structures, while regimes 1 and
2 can be viewed as special realizations of regime 3. In particular, we
notice that the latter is expected to represent also the most plausible
realization in real systems, in which both the gravitational and magnetic
fields contribute to determine the profiles of the fluid fields.

Let us then discuss the case of plasmas belonging to regime 3. The number
density profile is prescribed according to Eqs. (\ref{nde}) and (\ref{R3}).
We notice that, thanks to the analytical relationships (\ref{d1}) and (\ref%
{d2}) and the orderings (\ref{basic})$-$(\ref{eees}), any function of $%
\left( \psi ,\Phi _{\mathrm{G}}\right) $ can be equivalently expressed in
terms of $\left( r,\theta \right) $. Because of this, the rhs of Eq. (\ref%
{nde}) becomes now a generic function of $\left( r,\theta \right) $, namely
of the form $n_{\mathrm{s}}=n_{\mathrm{s}}\left( r,\theta \right) $. This
represents the most general kind of spatial dependence which is admitted by
the kinetic equilibrium. Hence, in the general case and in the absence of
other particular restrictions, suitable prescriptions of $n_{\mathrm{s}%
}\left( r,\theta \right) $ can be determined for regime 3 plasmas, according
to the real system to be studied, which admit maxima out of the equatorial
plane. This conclusion has a general character of validity and assures the
consistency of the kinetic theory presented here for collisionless
axisymmetric plasmas with the possible occurrence of levitating tori in the
external gravitational and magnetic fields of the type prescribed above.

We can now explore in more detail the present conclusion by considering
explicitly two possible physical realizations of this type of solution:

\textit{Case A:}\ In this first example we assume that both $\eta _{\mathrm{s%
}}$ and $T_{\mathrm{s}}$ in Eq. (\ref{nde}) are constant. From the physical
point of view, the requirement $\eta _{\mathrm{s}}=const.$ means that the
spatial variations of the number density are uniquely determined by the
exponential term (Maxwellian factor), which in turn depends on the
leading-order plasma temperature as well as on the rotational frequency,
gravitational potential and magnetic flux $\psi $. This choice is consistent
with previous literature (see for example \cite{AGN,Miller97,Miller2001}).

Concerning the condition $T_{\mathrm{s}}=const.$, this corresponds to a
leading-order plasma isothermal profile that is consistent with the kinetic
constraints that characterize the solution (see Section 7). We remark that
in the present framework, the isothermal condition can only be satisfied to
the leading-order, while for non-uniform plasmas, the full temperature
profile is generally non-constant because of higher-order kinetic effects.
These issues will be discussed in detail in Sections 7 and 8. In validity of
the prescription of constant $\eta _{\mathrm{s}}$ and $T_{\mathrm{s}}$, the
only freedom left concerns the functional dependence of $\Omega _{\mathrm{s}%
} $, which is considered of the form (\ref{R3}).

Under this assumption, the number density still remains of the type $n_{%
\mathrm{s}}\left( r,\theta \right) $. In this situation, it is convenient to
prescribe $n_{\mathrm{s}}\left( r,\theta \right) $ consistent with the
requirement of exhibiting maxima out of the equatorial plane, independently
of its actual representation given by Eq. (\ref{nde}). The prescription of a
physically-acceptable profile of $n_{\mathrm{s}}\left( r,\theta \right) $
must be done in such a way to reproduce observational or experimental data.
Once the profile of $n_{\mathrm{s}}\left( r,\theta \right) $ is set, since
also $\eta _{\mathrm{s}}$ and $T_{\mathrm{s}}$ are constant in this example,
then Eq. (\ref{nde}) can be inverted and used to uniquely derive the
expression of the corresponding species angular frequency $\Omega _{\mathrm{s%
}}=\Omega _{\mathrm{s}}\left( r,\theta \right) $ which determines the
levitating structure. In particular, the latter is obtained by solving the
quadratic algebraic equation%
\begin{equation}
\frac{M_{\mathrm{s}}}{2}R^{2}\Omega _{\mathrm{s}}^{2}+\frac{Z_{\mathrm{s}}e}{%
c}\psi \Omega _{\mathrm{s}}-M_{\mathrm{s}}\Phi _{\mathrm{G}}-T_{\mathrm{s}%
}\ln \frac{n_{\mathrm{s}}}{\eta _{\mathrm{s}}}=0.  \label{om}
\end{equation}

Hence, under these conditions, it is possible to introduce a density profile
which is in agreement with physical configurations and the existence of
off-equatorial tori. For leading-order isothermal systems this also
prescribes the form of the corresponding plasma rotation frequency according
to Eq. (\ref{om}). The extension of this solution method to the case of a
non-isothermal plasma species recquires the additional prescription of the
temperature profile, namely $T_{\mathrm{s}}=T_{\mathrm{s}}\left( r,\theta
\right) $, while the frequency $\Omega _{\mathrm{s}}$ can still be obtained
from Eq. (\ref{om}). In both cases we notice that the kinetic equilibrium
thus determined generally allows for the existence of two separate roots for 
$\Omega _{\mathrm{s}}$. If both are real, they should correspond to two
different admissible equilibria with opposite direction of plasma rotation
with respect to the external dipolar magnetic field orientation.

\textit{Case B:} As a second example, we assume the validity of the kinetic
constraint Eq. (\ref{R3}) for all the three structure functions. In
particular, here we treat the situation in which the condition%
\begin{equation}
n_{\mathrm{s}}\left( r,\theta \right) \equiv \eta _{\mathrm{s}}\left(
r,\theta \right)  \label{req1}
\end{equation}%
is identically satisfied in the configuration domain occupied by the
collisionless plasma species. From the physical point of view, Eq. (\ref%
{req1}) means that the number density profile $n_{\mathrm{s}}$ is not
modified by the Maxwellian exponential factor and coincides with the
pseudo-density $\eta _{\mathrm{s}}\left( r,\theta \right) $. This
requirement is satisfied when the exponential factor in Eq. (\ref{nde}) is
one. Hence, this condition is met for the species angular frequency
satisfying the algebraic quadratic equation%
\begin{equation}
\frac{M_{\mathrm{s}}}{2}R^{2}\Omega _{\mathrm{s}}^{2}+\frac{Z_{\mathrm{s}}e}{%
c}\psi \Omega _{\mathrm{s}}-M_{\mathrm{s}}\Phi _{\mathrm{G}}=0.  \label{omg}
\end{equation}

We notice that again Eq. (\ref{omg}) generates two roots for the frequency $%
\Omega _{\mathrm{s}}$, as for the case A discussed above. In addition, Eq. (%
\ref{omg}) holds for both isothermal and non-isothermal plasmas. Finally, in
validity of the $\sigma _{\mathrm{s}}-$ordering and the ordering (\ref{eees}%
), when $\ln \frac{n_{\mathrm{s}}}{\eta _{\mathrm{s}}}\sim O\left( 1\right) $%
, one can infer that the solutions of $\Omega _{\mathrm{s}}$ from Eq. (\ref%
{omg}) are asymptotically close to\ those from Eq. (\ref{om}), although the
number density and the temperature are not necessarily so.

To conclude this section, it is useful to make a qualitative comparison of
the results obtained here with those presented in Ref. \cite{JK1}, where the
existence of off-equatorial structures has been proved on the basis of a
fluid non-ideal MHD description. This involves in particular the inspection
of Eq. (41) for the pressure profile given in Ref. \cite{JK1}, which can
give rise to off-equatorial maxima for suitable choices of the coefficients
entering the same equation (see discussions in sections 3 and 4 in the same
reference). Indeed, pressure and density profiles are proportional (at least
to the leading-order) when the condition $T_{\mathrm{s}}=const.$ applies
(see also Eq. (\ref{ptot}) below). In such a case, it is immediate to verify
that the rhs of Eq. (41) in Ref. \cite{JK1} can be effectively expressed as
a function of $\psi $ and $\Phi _{\mathrm{G}}$ only, in agreement with the
prescription holding for regime 3 plasmas. Although the two treatments
(i.e., the present one and Ref. \cite{JK1}) consider different physical
conditions for the levitating plasma, the consistency pointed out here
establishes a notable result. In fact, first it shows that, as anticipated
in the Introduction, the kinetic theory developed in this paper and its
analytical formulation allow for direct comparisons with previous literature
works based on fluid approaches. Second, it proves that fluid results can in
principle be reproduced consistently on the basis of a kinetic treatment,
thus extending their validity to a wider class of plasma regimes. Third, in
turn it reinforces the statement given above concerning the general
character of the present kinetic theory for regime 3 plasmas in providing a
suitable mathematical and physical framework for the investigation of
off-equatorial structures.

\section{Kinetic corrections}

In this section we provide the explicit representation of the kinetic
corrections $h_{\mathrm{s}}^{\left( 1\right) }$, $h_{\mathrm{s}}^{\left(
2\right) }$ and $h_{\mathrm{s}}^{\left( 3\right) }$ introduced in the
Chapman-Enskog representation of the equilibrium KDF given by Eq. (\ref{CE}%
). The inclusion of these contributions is necessary for the complete
solution of the equilibrium problem in the framework of Vlasov-Maxwell
description of collisionless plasmas. In particular, these terms represent
the deviations of the KDF from a Maxwellian distribution and arise because
of the constraints imposed by single-particle phase-space conservation laws
on the solution itself. The precise definition of the first-order terms $h_{%
\mathrm{s}}^{\left( 1\right) }$, $h_{\mathrm{s}}^{\left( 2\right) }$ and $h_{%
\mathrm{s}}^{\left( 3\right) }$ is also required to distinguish the solution
among the three kinetic regimes pointed out above.

In detail, the first-order corrections $h_{\mathrm{s}}^{\left( 1\right) }$
and $h_{\mathrm{s}}^{\left( 2\right) }$ originate from the perturbative
treatment of the implicit phase-space dependences carried by the
structure-functions $\Lambda _{\ast \mathrm{s}}$ entering the equilibrium
KDF $f_{\ast \mathrm{s}}$. They are found to be given by:%
\begin{eqnarray}
h_{\mathrm{s}}^{\left( 1\right) } &\equiv &\frac{cM_{\mathrm{s}}}{Z_{\mathrm{%
s}}e}R\left[ A_{1\mathrm{s}}+\frac{p_{\varphi \mathrm{s}}\Omega _{\mathrm{s}}%
}{T_{\mathrm{s}}}A_{2\mathrm{s}}\right] v_{\varphi }  \nonumber \\
&&+\frac{cM_{\mathrm{s}}}{Z_{\mathrm{s}}e}R\left[ \frac{E_{\mathrm{s}%
}-\Omega _{\mathrm{s}}p_{\varphi \mathrm{s}}}{T_{\mathrm{s}}}-\frac{3}{2}%
\right] A_{3\mathrm{s}}v_{\varphi },  \label{h1}
\end{eqnarray}%
\begin{eqnarray}
h_{\mathrm{s}}^{\left( 2\right) } &\equiv &\left[ \frac{E_{\mathrm{s}%
}-\Omega _{\mathrm{s}}p_{\varphi \mathrm{s}}}{T_{\mathrm{s}}}-\frac{3}{2}%
\right] C_{3\mathrm{s}}\left( \frac{1}{2}v^{2}+\frac{{Z_{\mathrm{s}}e}}{M_{%
\mathrm{s}}}\Phi \right)  \nonumber \\
&&+\left[ C_{1\mathrm{s}}+\frac{p_{\varphi \mathrm{s}}\Omega _{\mathrm{s}}}{%
T_{\mathrm{s}}}C_{2\mathrm{s}}\right] \left( \frac{1}{2}v^{2}+\frac{{Z_{%
\mathrm{s}}e}}{M_{\mathrm{s}}}\Phi \right) ,  \label{h2}
\end{eqnarray}%
where the following definitions have been introduced:%
\begin{eqnarray}
A_{1\mathrm{s}} &\equiv &\frac{\partial \ln \eta _{\mathrm{s}}}{\partial
\psi },A_{2\mathrm{s}}\equiv \frac{\partial \ln \Omega _{\mathrm{s}}}{%
\partial \psi },A_{3\mathrm{s}}\equiv \frac{\partial \ln T_{\mathrm{s}}}{%
\partial \psi }, \\
C_{1\mathrm{s}} &\equiv &\frac{\partial \ln \eta _{\mathrm{s}}}{\partial
\Phi _{\mathrm{G}}},C_{2\mathrm{s}}\equiv \frac{\partial \ln \Omega _{%
\mathrm{s}}}{\partial \Phi _{\mathrm{G}}},C_{3\mathrm{s}}\equiv \frac{%
\partial \ln T_{\mathrm{s}}}{\partial \Phi _{\mathrm{G}}}.
\end{eqnarray}%
Hence, $h_{\mathrm{s}}^{\left( 1\right) }$ and $h_{\mathrm{s}}^{\left(
2\right) }$ are polynomial functions of the particle velocity which contain
diamagnetic and energy-correction contributions and depend on the so-called
thermodynamic forces $\frac{\partial \Lambda _{\mathrm{s}}}{\partial \psi }$
and $\frac{\partial \Lambda _{\mathrm{s}}}{\partial \Phi _{\mathrm{G}}}$.
The latter represent the gradients of the structure functions across
equipotential magnetic and gravitational surfaces and arise in collisionless
plasmas characterized by non-uniform fluid fields. Consistency of these
expressions with the $\varepsilon _{\mathrm{s}}$ and $\sigma _{\mathrm{s}}$
ordering assumptions requires that%
\begin{equation}
\frac{cM_{\mathrm{s}}}{Z_{\mathrm{s}}e}R\left[ \left( \frac{E_{\mathrm{s}%
}-\Omega _{\mathrm{s}}p_{\varphi \mathrm{s}}}{T_{\mathrm{s}}}-\frac{3}{2}%
\right) A_{3\mathrm{s}}\right] v_{\varphi }\lesssim O\left( \varepsilon _{%
\mathrm{s}}\right) ,
\end{equation}%
\begin{equation}
\left[ \frac{E_{\mathrm{s}}-\Omega _{\mathrm{s}}p_{\varphi \mathrm{s}}}{T_{%
\mathrm{s}}}-\frac{3}{2}\right] C_{3\mathrm{s}}\left( \frac{1}{2}v^{2}+\frac{%
{Z_{\mathrm{s}}e}}{M_{\mathrm{s}}}\Phi \right) \lesssim O\left( \sigma _{%
\mathrm{s}}\right) ,
\end{equation}%
which implies that $T_{\mathrm{s}}$ must actually be of the form $T_{\mathrm{%
s}}=T_{\mathrm{s}}\left( \varepsilon _{\mathrm{s}}^{k}\psi ,\sigma _{\mathrm{%
s}}^{k}\Phi _{\mathrm{G}}\right) $, with $k\geq 1$, i.e. at most
slowly-dependent on $\psi $ and $\Phi _{\mathrm{G}}$. This conclusion
motivates the choice done in Section 6 to treat isothermal plasmas (to
leading-order). As pointed out above, the contribution $h_{\mathrm{s}%
}^{\left( 1\right) }$ is null for gravitationally-bound plasmas, while $h_{%
\mathrm{s}}^{\left( 2\right) }$ vanishes for magnetized plasmas. Instead, both terms are
present in the equilibrium solution for plasmas belonging to regime 3. We
also notice that the $\sigma _{\mathrm{s}}-$expansion generates
contributions in $h_{\mathrm{s}}^{\left( 2\right) }$ which are proportional
to both particle kinetic energy and ES potential. In particular, terms which
depend on $\Phi $ contribute to the occurrence of ES corrections to the
kinetic solution and the corresponding fluid fields.

Finally, the last contribution $h_{\mathrm{s}}^{\left( 3\right) }$
originates from the validity of the $\lambda -$ordering (\ref{eees}) when
this is taken into account in the expression for the leading-order number
density, and results in the following term%
\begin{equation}
h_{\mathrm{s}}^{\left( 3\right) }\equiv \frac{Z_{\mathrm{s}}e\Phi }{M_{%
\mathrm{s}}\Phi _{\mathrm{G}}}.  \label{enne-esse}
\end{equation}%
It must be stressed that the ES contributions arising in $h_{\mathrm{s}%
}^{\left( 2\right) }$ and $h_{\mathrm{s}}^{\left( 3\right) }$ originate from
different perturbative treatments. In fact, $h_{\mathrm{s}}^{\left( 3\right)
}$ follows from the $\lambda -$ordering and is common to all the regimes
considered here when Eq. (\ref{eees}) applies. Instead the terms in $h_{%
\mathrm{s}}^{\left( 2\right) }$ can only be included when the $\sigma _{s}-$%
ordering applies (regimes 1 and 3).

To conclude the section, it is worth pointing out that the treatment of the
first-order kinetic corrections displayed here requires the following
preliminary steps:\newline
1) The precise identification of the appropriate plasma collisionless
kinetic regime.\newline
2)\ The prescription of the leading-order spatial profiles of the structure
functions, consistent with the kinetic constraints for each regime.\newline
3)\ The evaluation of the thermodynamic forces and the explicit calculation
of the ES potential.\newline
In particular, the existence of the equilibria determined here is subject to
the validity of the Maxwell equations, i.e. the Poisson equation for the ES
potential $\Phi $ and Ampere's equation (see Section 9).

\section{Equation of state}

In this section we proceed with the calculation of the EoS corresponding to the kinetic equilibrium determined here. This
requires in particular to compute the species pressure tensor $\underline{%
\underline{\Pi }}_{\mathrm{s}}$ carried by the KDF $f_{\ast \mathrm{s}}$ and
defined as%
\begin{equation}
\underline{\underline{\Pi }}_{\mathrm{s}}\equiv M_{\mathrm{s}}\int_{\Gamma _{%
\mathbf{v}}}\mathrm{{d}\mathbf{v}\left( \mathbf{v}-\mathbf{V}_{{s}}\right) \left( 
\mathbf{v}-\mathbf{V}_{{s}}\right) f_{\ast {s},}}  \label{pten}
\end{equation}%
where $\Gamma _{\mathbf{v}}$ denotes the velocity domain of integration.
Since $f_{\ast \mathrm{s}}$ is isotropic with respect to quadratic particle
velocity dependences, one can immediately infer that each species pressure
tensor is isotropic and of the form $\underline{\underline{\Pi }}_{\mathrm{s}%
}=p_{\mathrm{s}}^{\mathrm{tot}}\underline{\underline{\mathbf{I}}}$, where $%
p_{\mathrm{s}}^{\mathrm{tot}}=n_{\mathrm{s}}^{\mathrm{tot}}T_{\mathrm{s}}^{%
\mathrm{tot}}$ denotes the thermal scalar pressure, with $n_{\mathrm{s}}^{%
\mathrm{tot}}$ and $T_{\mathrm{s}}^{\mathrm{tot}}$ being respectively the
species total number density and temperature associated with $f_{\ast 
\mathrm{s}}$. The calculation of $p_{\mathrm{s}}^{\mathrm{tot}}$ can be
carried out analytically for thermal particles when the Chapman-Enskog
representation (\ref{CE}) applies. In the following we consider this case.
Furthermore, consistent with the $\varepsilon _{\mathrm{s}}$ and $\sigma _{%
\mathrm{s}}$ orderings, in the first-order terms $h_{\mathrm{s}}^{\left(
1\right) }$ and $h_{\mathrm{s}}^{\left( 2\right) }$ we approximate the
canonical momentum $p_{\varphi \mathrm{s}}$ and the energy $E_{\mathrm{s}}$
respectively with $\frac{Z_{\mathrm{s}}e}{c}\psi $ and $M_{\mathrm{s}}\Phi _{%
\mathrm{G}}$. Hence, under these assumptions, one can prove that the scalar
pressure can be represented as%
\begin{equation}
p_{\mathrm{s}}^{\mathrm{tot}}=n_{\mathrm{s}}T_{\mathrm{s}}\left[ 1+\sigma _{%
\mathrm{s}}\Delta p_{\mathrm{s}}^{\left( 2\right) }+\lambda h_{\mathrm{s}%
}^{\left( 3\right) }\right] ,  \label{ptot}
\end{equation}%
where $p_{\mathrm{s}}\equiv n_{\mathrm{s}}T_{\mathrm{s}}$ is the
leading-order term, with $n_{\mathrm{s}}$ being defined by Eq. (\ref{nde}).
In addition we notice that the term $h_{\mathrm{s}}^{\left( 1\right) }$
associated with the $\varepsilon _{\mathrm{s}}-$expansion does not
contribute to the EoS because it is odd in the azimuthal component of
particle velocity. Instead, $h_{\mathrm{s}}^{\left( 3\right) }$ does not
depend explicitly on particle velocity and\ therefore it is not affected
when the integral (\ref{pten}) is computed on $\Gamma _{\mathbf{v}}$. Hence,
its contribution in Eq. (\ref{ptot}) is simply proportional to $p_{\mathrm{s}%
}$ and represents part of the ES corrections which enter the definition of
the total pressure $p_{\mathrm{s}}^{\mathrm{tot}}$. Finally, invoking Eq. (%
\ref{h2}), explicit calculation gives for $\Delta p_{\mathrm{s}}^{\left(
2\right) }$ the following result:%
\begin{equation}
\Delta p_{\mathrm{s}}^{\left( 2\right) }=\left( 2\frac{{Z_{\mathrm{s}}e}}{M_{%
\mathrm{s}}}\Phi +4\frac{T_{\mathrm{s}}}{M_{\mathrm{s}}}\right) Y_{\mathrm{s}%
},
\end{equation}%
where%
\begin{equation}
Y_{\mathrm{s}}\equiv C_{1\mathrm{s}}+\frac{\frac{Z_{\mathrm{s}}e}{c}\psi
\Omega _{\mathrm{s}}}{T_{\mathrm{s}}}C_{2\mathrm{s}}+\left( \frac{M_{\mathrm{%
s}}\Phi _{\mathrm{G}}-\Omega _{\mathrm{s}}\frac{Z_{\mathrm{s}}e}{c}\psi }{T_{%
\mathrm{s}}}-\frac{3}{2}\right) C_{3\mathrm{s}}.
\end{equation}

From this result it is interesting to point out that, although to the
leading-order the species pressure coincides with the thermal pressure, the
first-order corrections introduce deviations in the EoS that are distinctive
for collisionless plasmas. In particular, the energy-correction
contributions which enter through $\Delta p_{\mathrm{s}}^{\left( 2\right) }$
are associated with the gradients of structure functions across
gravitational equipotential surfaces and include also ES corrections
proportional to $\Phi $. These terms however vanish in uniform collisionless
plasmas. On the other hand, the ES correction to the EoS carried by $h_{%
\mathrm{s}}^{\left( 3\right) }$ is independent and follows uniquely from the 
$\lambda -$ordering introduced above between ES and gravitational potential
energy. Clearly, all the first-order contributions in the EoS arise as
specifically-kinetic effects, which characterize the kinetic treatment of
collisionless plasmas.

\section{The Maxwell equations}

In this section we analyze the constraints which are posed by the Maxwell
equations on the kinetic equilibria. These concern in particular the
validity of the orderings (\ref{basic}) and (\ref{eees}) and for this reason
they apply to all the plasma regimes identified above.

We consider first the Poisson equation for the ES potential, which is
written as%
\begin{equation}
\nabla ^{2}\Phi =-4\pi \sum_{\mathrm{s}}Z_{\mathrm{s}}en_{\mathrm{s}}^{%
\mathrm{tot}}.  \label{Po1}
\end{equation}%
In the general case the solution is non-trivial, because the total number
density $n_{s}^{tot}$ depends both implicitly and explicitly on the ES
potential itself (see for example \cite{Cr2011a,Cr2011}). However, the
solution simplifies in validity of the sub-ordering expansion (\ref{eees})
introduced above. In fact in this case the ES potential enters the kinetic
solution only through the first-order corrections to the equilibrium KDF.
Therefore, consistent with the orderings introduced in the present work and
the perturbative theory developed here, one can obtain an asymptotic
solution for $\Phi $ by considering only the leading-order contribution to
the species number density. Thus, when the said sub-ordering applies,
neglecting corrections of $O\left( \sigma _{\mathrm{s}}\right) $ and $%
O\left( \zeta _{\mathrm{s}}\right) $ and invoking Eq. (\ref{nde}), the
Poisson equation becomes to this accuracy:%
\begin{equation}
\nabla ^{2}\Phi =S\left( \mathbf{x}\right) ,  \label{Po2}
\end{equation}%
where the source term $S\left( x\right) $ is defined as%
\begin{equation}
S\left( \mathbf{x}\right) \equiv -4\pi \sum_{\mathrm{s}}Z_{\mathrm{s}}e\eta _{\mathrm{%
s}}\exp \left[ \frac{\frac{M_{\mathrm{s}}}{2}R^{2}\Omega _{\mathrm{s}}^{2}+%
\frac{Z_{\mathrm{s}}e}{c}\psi \Omega _{\mathrm{s}}-M_{\mathrm{s}}\Phi _{%
\mathrm{G}}}{T_{\mathrm{s}}}\right] .  \label{tesader}
\end{equation}%
Here $S\left( \mathbf{x}\right) $ does not depend on $\Phi $, and therefore the ES
potential can be readily obtained by integrating Eq. (\ref{Po2}) yielding%
\begin{equation}
\Phi \left( \mathbf{x}\right) =\int \mathrm{{d}\mathbf{x}^{\prime }G\left(
\mathbf{x}-\mathbf{x}^{\prime }\right) S\left( \mathbf{x}^{\prime }\right) ,}
\end{equation}%
with $G\left( \mathbf{x}-\mathbf{x}^{\prime }\right) $ being the corresponding Green function.
For the consistency of the theory, the solution for $\Phi $ given by the
previous equation must be checked \textquotedblleft a
posteriori\textquotedblright\ to satisfy the initial ordering (\ref{eees}).
In particular, this can represent a constraint condition for the magnitude of the species number densities which contribute to the ES
potential through the system charge density (\ref{tesader}). Manifestly, the
validity of the ordering (\ref{eees}) is necessary for the present theory to
apply, and for this reason the calculation of $\Phi $ represents the
ultimate step to be done in order to warrant the consistency of the
treatment.

Similar considerations apply to the Ampere equation, which determines the
self-generation of magnetic field by the equilibrium collisionless plasma.
The Ampere equation is written as%
\begin{equation}
\nabla \times \mathbf{B}^{\mathrm{self}}=\frac{4\pi }{c}\mathbf{J}^{\mathrm{%
tot}},  \label{Ampere-2}
\end{equation}%
where $\mathbf{B}^{\mathrm{self}}$ is defined in Eq. (\ref{bself}) and $%
\mathbf{J}^{\mathrm{tot}}$ is the total current density, which is given by%
\begin{equation}
\mathbf{J}^{\mathrm{tot}}\equiv \sum\limits_{\mathrm{s}}\mathbf{J}_{\mathrm{s%
}}^{\mathrm{tot}}=\sum\limits_{\mathrm{s}}Z_{\mathrm{s}}en_{\mathrm{s}}^{%
\mathrm{tot}}\mathbf{V}_{\mathrm{s}}^{\mathrm{tot}},
\end{equation}%
with $\mathbf{V}_{\mathrm{s}}^{\mathrm{tot}}$ being the species flow
velocity. It is immediate to prove that, in the present formulation, $%
\mathbf{V}_{\mathrm{s}}^{\mathrm{tot}}$ is purely azimuthal at equilibrium;
in fact additional components of the velocity can only arise in the presence
of temperature anisotropy, see for example \citet{Cr2010,Cr2011a,Cr2011}.
Again, consistent with the perturbative treatment presented here, one can
retain only the leading-order contributions to $\mathbf{J}^{\mathrm{tot}}$
in Eq. (\ref{Ampere-2}). Under this assumption Eq. (\ref{Ampere-2}) becomes%
\begin{equation}
\nabla \times \mathbf{B}^{\mathrm{self}}=\frac{4\pi }{c}\sum\limits_{\mathrm{%
s}}Z_{\mathrm{s}}en_{\mathrm{s}}\mathbf{V}_{\mathrm{s}},  \label{GSS}
\end{equation}%
where $n_{\mathrm{s}}$ is given by Eq. (\ref{nde}) and $\mathbf{V}_{\mathrm{s%
}}=R\Omega _{\mathrm{s}}\mathbf{e}_{\varphi }$ (see Eq. (\ref{fm})). Eq. (%
\ref{GSS}) represents a generalized Grad-Shafranov equation for the poloidal
magnetic flux $\psi _{\mathrm{self}}$ in which the source term on the rhs
depends only on explicitly known quantities. The solution for $\mathbf{B}^{%
\mathrm{self}}$ which results from Eq. (\ref{GSS}) must then be checked
\textquotedblleft a posteriori\textquotedblright\ to verify the ordering
condition (\ref{basic}) introduced above, which is necessary in
order to warrant the validity of the theory and its analytical development.
In this case Eq. (\ref{basic}) can represent a constraint for the magnitude of the species rotation angular frequencies which contribute to
the system charge current.

\section{Conclusions}

In this paper, a theoretical investigation of equilibrium configurations for
collisionless non-relativistic and axisymmetric plasmas has been presented,
taking into account the role of a central spherically-symmetric
gravitational field. The formulation is based on a multi-species kinetic
approach developed in the framework of the Vlasov-Maxwell description. The
case of astrophysical plasmas arising in the gravitational field of compact
objects and in the presence of both an external dipolar magnetic field and
self electromagnetic fields has been treated.

Three different plasma regimes have been identified which are characetrized
by distinctive kinetic orderings. It has been proved that in all cases
consistent kinetic equilibria can be determined, with the kinetic
distribution function being expressed in terms of generalized Maxwellian
functions. It has been shown that the three regimes differ by the form of
the kinetic constraints which are imposed on the equilibrium solutions and
which uniquely follow from phase-space single-particle conservation laws.

By imposing appropriate orderings on the self electromagnetic fields and by
developing a suitable perturbative theory, an analytical treatment of the
equilibria has been proposed. In terms of this, several issues have been
addressed. First, the conditions of existence of equilibrium structures
corresponding to off-equatorial tori have been investigated. It has been
shown that these systems can generally arise for the regime which has been
referred to here as magnetized and gravitationally-bound plasmas. This
analysis can be important from the astrophysical point of view, since
off-equatorial tori may represent a physically-realizable model of
magnetized coronal plasmas which are believed to characterize accretion
discs. In addition, the treatment based on kinetic theory can pose the basis
for comparison with analogous fluid results carried out in terms of MHD
theory.

As a second application, the plasma equation of state has been determined
analytically and expressed in terms of the pressure tensor. It has been
shown that the latter exhibits deviations from the thermal pressure
characteristic of collisional plasmas because of the existence of
specifically-kinetic effects. These have been identified with diamagnetic,
energy-correction and electrostatic contributions which apply in combination
with the occurrence of non-uniform fluid fields.

Finally, the validity of the Poisson and Ampere equations have been
addressed, showing that they can introduce non-trivial constraints on the
magnitude of the plasma number density and flow velocity for the consistency
with the orderings introduced in the theory developed here.

The conclusions established in this work can be relevant for future
investigations of astrophysical plasmas in equilibrium configurations, with
particular focus on collisionless plasmas in accretion discs and
off-equatorial tori associated with compact objects.

\acknowledgments

Financial support by the Italian Foundation \textquotedblleft Angelo Della
Riccia\textquotedblright\ (Firenze, Italy) is acknowledged by C.C. C.C., J.K., P.S. and Z.S. would like to express their acknowledgment for the Institutional
support of Faculty of Philosophy and Science, Silesian University in Opava
(Czech Republic) and thank the project Synergy CZ.1.07/2.3.00/20.0071
sustaining the international collaboration. V.K. thanks Czech Science
Foundation GA\v{C}R 13-00070J.






\clearpage

\end{document}